\documentclass[useAMS,usegraphicx]{mn2e}
\usepackage{amssymb}
\usepackage{lscape}
\usepackage{times}
\usepackage{pifont}
\usepackage{amsfonts}
\usepackage{amsmath}
\usepackage{cleveref}							% clever cross referencing
\usepackage{url}		
\usepackage{rotating}
\usepackage{stfloats}
\usepackage{longtable,lscape}
\usepackage{tabularx}

\def\kms{km${\rm s}^{-1}$}
\def\arcdeg{\hbox{$^\circ$}}
\def\arcmin{\hbox{$^\prime$}}
\def\arcsec{\hbox{$^{\prime\prime}$}}
\def\ergcms{erg\,cm$^{-2}$s$^{-1}$}
\def\micron{$\mu$m}
\def\ha{H$\alpha$}
\def\hb{H$\beta$}

\def\hd{H$\delta$}
\def\NII{[N\,\textsc{ii}]}

\def\OIII{[O\,\textsc{iii}]}
\def\OIV{[O\,\textsc{iv}]}

\def\HeII{He\,\textsc{ii}}

\def\p0{\phantom{0}}

\def\lessim{\raise-.5ex\hbox{$\buildrel<\over{\scriptstyle\mathtt{\sim}}$}}
\def\grtsim{\raise-.5ex\hbox{$\buildrel>\over{\scriptstyle\mathtt{\sim}}$}}

\title[DT\,Serpentis]{DT Serpentis: neither a symbiotic star nor a planetary nebula associate}
\author[D.J. Frew et al.]{David J. Frew$^{1,2}$\thanks{E-mail.david.frew@mq.edu.au}, Joao Bento$^{1,2}$, 
Ivan S. Boji\v{c}i\'c$^{1,2,3}$ and Quentin A. Parker$^{1,2,3}$ \\
$^{1}$Department of Physics and Astronomy, Macquarie University, Sydney, NSW 2109, Australia\\
$^{2}$Research Centre in Astronomy, Astrophysics \& Astrophotonics, Macquarie University, Sydney, NSW 2109, Australia\\
$^{3}$Australian Astronomical Observatory, PO Box 915, North Ryde, NSW 1670, Australia
}

\begin{document}

\date{Accepted ; Received ; in original form }
\pagerange{\pageref{firstpage}--\pageref{lastpage}} \pubyear{}

\maketitle
\label{firstpage}

\begin{abstract}
We  present an alternative interpretation for the putative symbiotic star DT Serpentis, and its proposed planetary nebula (PN), recently announced by Munari et al.   Our analysis is  based on their data combined with additional archival data trawled from Virtual Observatory databases.  We show that the star known as DT\,Ser is not a symbiotic star, and is merely superposed on the newly discovered but unrelated background PN. There is no evidence for any periodic variability for DT\,Ser as expected for a symbiotic star. We further establish that there is no physical association between DT\,Ser and the PN, which has a considerably higher extinction, befitting the larger distance we estimate.  The significantly different radial velocities of the star and nebula also likely preclude any association.  Finally, we show that the mid-infrared source detected by the IRAS and WISE surveys is actually coincident with the PN so there is no evidence for DT\,Ser being a dusty post-AGB star. 
\end{abstract}

\begin{keywords}
planetary nebulae. general -- planetary nebulae. individual -- symbiotic stars -- surveys
\end{keywords}

\section{Introduction}

In this paper we suggest an alternative interpretation for the proposed association between a newly discovered planetary nebula (PN) and the ``known'' symbiotic star DT\,Serpentis recently announced by Munari et al. (2013, hereafter MC13).   As discussed in their paper, the apparent connection of a symbiotic star in a round planetary nebula, closely associated with a potential post-AGB star, would be a unique occurrence, and this led us to more closely investigate this very interesting association.  Our new interpretation is based solely on the data presented by MC13, combined with multi-wavelength data trawled from Virtual Observatory databases.  
As the new PN does not appear to have a common name, we designate it as MCS\,1, after the first three authors of MC13.   This paper is arranged as follows:  in \S\,\ref{sec:data} we describe the archival data 
used, and in \S\,\ref{sec:DTSer} we assess the nature of DT\,Ser and the evidence for its variability.  We provide a discussion of the PN, its distance and properties in \S\,\ref{sec:PN}, and present our conclusions in \S\,\ref{sec:summary}.

%%%%%%%%%%%%%%%%%%%%%%%%%%%%%%%%%%%%%%%%%%%%%%%%%%%%%%%%%%%%%%%%%%%%%%%%%%%%%%%%%%%%%%%%%%%%%%%%%%%%%
%%%%%%%%%%%%%%%%%%%%%%%%%%%%%%%%%%%%%%%%%%%%%%%%%%%%%%%%%%%%%%%%%%%%%%%%%%%%%%%%%%%%%%%%%%%%%%%%%%%%%
\section{Archival Data}\label{sec:data}

To supplement the spectroscopic and photometric data presented by MC13, we searched for archival multi-wavelength data for DT Ser and MCS\,1 using a range of online tools (e.g. Frew et al. 2011).  We interrogated the Aladin Sky Atlas, the SIMBAD database, the Vizier service\footnote{Aladin, SIMBAD and Vizier are accessible from the Centre de Donn\'ees Astronomiques (CDS) at \url{http://cdsweb.u-strasbg.fr/}},  the SkyView Virtual Observatory\footnote{\url{http.//skyview.gsfc.nasa.gov/}}, the NASA/IPAC Infrared Science Archive\footnote{\url{http://irsa.ipac.caltech.edu/}}, and SkyDOT\footnote{\url{http://skydot.lanl.gov/}} (Sky Database for Objects in Time-Domain).
Aperture photometry of the  low resolution narrowband \ha\ (+ \NII) image from  the Southern H-Alpha Sky Survey Atlas (SHASSA) (Gaustad et al. 2001) provided the integrated \ha\ flux of the nebula.   Multi-wavelength broadband images and data were also obtained from the Galaxy Evolution Explorer (GALEX) survey (Morrissey et al. 2007), the SuperCOSMOS Sky Survey (SSS; Hambly et al. 2001), the Two Micron All Sky Survey (2MASS; Skrutskie et al. 2006), and the Wide-Field Infrared Survey Explorer (WISE; Wright et al. 2010) Survey.   Additional photometry for DT\,Ser were retrieved from the VizieR service, or the literature where noted.  

%%%%%%%%%%%%%%%%%%%%%%%%%%%%%%%%%%%%%%%%%%%%%%%%%%%%%%%%%%%%%%%%%%%%%%%%%%%%%%%%%%%%%%%%%%%%%%%%%%%%%%%%%%%

\section{Is DT\,Ser a symbiotic star?}\label{sec:DTSer}

Symbiotic stars are interacting binary systems, in which an evolved star (usually a red giant) transfers matter onto a hotter, more compact companion star, most typically a white dwarf (Kenyon 1986). Depending on the relative contributions of the cool star and any dust enveloping the system, they are subdivided into S-type (stellar) and D-type (dusty) systems.  A small group of systems containing hotter companion stars (spectral types of F, G and K) are the so-called yellow symbiotic stars; a subset of these showing thermal emission from hot dust are designated D$^{\prime}$-type systems (Allen 1982; Belczy\'nski et al. 2000).  

Symbiotic stars can show a complex range of photometric variations, depending on the exact orbital parameters and the nature, luminosity, and mass-loss rate of the secondary star.  Symbiotic stars are known to show eclipses, ellipsoidal variability or a reflection effect due to orbital motion (mostly S-type and D$^{\prime}$-type), pulsational variability of the cool component (D-type and S-type), dust obscuration events (D-type), short-period flickering associated with the accretion disk, high- and low-activity states, and rare nova-like outbursts originating from the white dwarf in the system (Miko\l ajewska 2001; Gromadzki et al. 2009; Gromadzki, Miko\l ajewska \& Soszy\'nski  2013; Angeloni et al. 2014).

\subsection{Historical Background}\label{sec:background}

DT\,Ser was originally discovered by Hoffmeister (1949) from photographic plates taken at Sonneberg Observatory, later appearing on the MVS charts (Hoffmeister 1957). The latest edition of the General Catalogue of Variable Stars (GCVS; Samus et al. 2007-2012) gives a photographic (blue) range of 13.2--13.9 mag, based on the analyses of G\"otz (1957) and Meinunger (1980).   %A CHECK OF THESE DETAILS NEEDED

Nebular emission  around this star was first noted by Bond (1978)  who wrote:  ``\emph{Superposed on an absorption spectrum of about type G0 are \OIII\ emission lines at 4959--5007\,\AA.  The object appears stellar on the Palomar Sky Survey and at the telescope, with a slightly fainter companion $\sim$5\arcsec\ away.}''  The PN was not noticed in any broad-band imagery, and on this basis Bond suggested the star was probably a symbiotic variable.  
We follow the nomenclature of MC13 in referring to DT\,Ser as star\,A and the fainter companion 5\arcsec\ away as star\,B, which is the ionizing star of MCS\,1.  

Cieslinski et al. (1997) however, suggested that the symbiotic star is the companion noticed by Bond (1978), in effect transferring the designation DT\,Ser to the central star, and categorising star\,A as a field star.  Henden \& Munari (2001)  concurred, further commenting that the old GCVS data refer to the combined light of the two stars.  Cieslinski et al. (1997) further suggested DT\,Ser was a yellow symbiotic star (Schmid \& Nussbaumer 1993; Jorissen et al. 2005), some of which are known to have resolved nebulae associated with them (e.g. Schwarz 1991; Van Winckel et al. 1995; Miszalski et al. 2012), so this interpretation seemed feasible.  Based on this information, DT\,Ser was included as a suspected symbiotic star in the catalogue of Belczy\'nski et al. (2000), who commented that the cool component might not be physically associated with the emission-line source.
In addition, Carballo, Wesselius \& Whittet (1992) and Kinnunen \& Skiff (2000) noted the coincidence of an IRAS mid-infrared source with DT\,Ser.
We will assess the nature of DT\,Ser and the likelihood of a physical association with MCS\,1 in the following sections.
 % ... and they suggested it was a yellow symbiotic star.

\subsection{Spectral Classification}\label{sec:spectrum}
MC13 determined the spectral type of DT\,Ser to be F8, with an indeterminate luminosity class.  The spectral type is consistent with the older determination of G0e by Bond (1978), however Cieslinski, Steiner \& Jablonski (1998) obtained a later spectral type of G2--K0\,III--I,  noting Balmer, \OIII, and \HeII\ lines in emission (this classification was adopted by M\"urset \& Schmid 1999).  We carefully examined enlarged reproductions of DT\,Ser's spectrum from Figure\,3 of MC13  and find the weakness of the G-band and the strength of the Balmer lines to be inconsistent with a type later than mid-G, and take at face value the F8 spectral type.  From the complete absence of the luminosity sensitive Sr\,II $\lambda$4077 line adjacent to \hd, and the $\lambda\lambda$4172--79\,\AA\ blend of CN, we can clearly exclude a supergiant or bright-giant luminosity (Gray \& Corbally 2009) though we cannot differentiate between the luminosity classes III, IV or V.  Furthermore, a luminosity class of III--V is also a better fit to the overall strength of the many neutral metal lines in the blue.\footnote{A simple guess \emph{a priori} is that the star is a main-sequence object, as F8 giants are located in the Hertzsprung Gap on the HR diagram and are relatively uncommon stars.} 

Indeed, using high-resolution Echelle spectra, MC13 note only a late-F continuum with no emission lines, and find the centroid of the Balmer, \OIII, and \HeII\ nebular emission lines to be offset from DT\,Ser.  Thus, the emission lines recorded by Bond (1978) and Cieslinski et al. (1997, 1998) are undoubtedly due to the small PN partly superposed on the image of DT\,Ser, especially at small plate scales. 
%{\bf CAN WE BE MORE SPECIFIC?}
On the basis of all available spectroscopic data, we assert that neither star\,A or star\,B is a bona fide symbiotic star, and that the observed nebular lines seen by earlier workers solely originate in the extended PN.   

\subsection{Photometry}\label{sec:photometry}

%the authors claim that they cannot detect any periodic behaviour symptomatic of a symbiotic star. They should, however, detail what kind of variations they expect, at which level, on which time scales and due to which physical mechanisms. 
% Q: what is the lowest amplitude symbiotic variability level so far uncovered?

We compiled all reliable survey photometry for DT\,Ser from the references given in Table\,\ref{table:phot}, including several reddening-corrected colour indices.  The $U-B$ colour index is the most useful luminosity diagnostic for F-type stars, as it measures the depth of the Balmer discontinuity which is sensitive to surface gravity (e.g. Bond 1997); supergiants have large Balmer jumps and redder $U-B$ colours.  DT\,Ser plots close to the main-sequence locus (and well away from the supergiant locus) in the $B-V$ versus $U-B$ colour-colour diagram (Johnson \& Morgan 1953), providing additional support for a luminosity class of III-V.

\begin{table}
{\footnotesize	
\begin{center}
\caption{Summary of photometric measurements for DT\,Ser.  The de-reddened colours were calculated assuming $E(B-V)$ = 0.25.}
\label{table:phot}
\begin{tabular}{cll}
\hline
Waveband~ 		&      ~~~~~~~mag 			&  Source		\\
\hline							
$U$                  	&  13.66	$\pm$ 0.01			&  HM01, HM08	 \\%
$B$                  	&  13.56 $\pm$ 0.10    		&  CS97			 \\%check uncertainty
$B$                  	&  13.55 $\pm$ 0.00			&  HM01, HM08	 \\%
$B$                  	&  13.57 $\pm$ 0.04     		& APASS		\\% n = 6, DR 7
$g'$                 	&  13.17 $\pm$ 0.05$^a$    	& APASS         	 \\
$V$                  	&  ~~12.8 $\pm$ 0.1    		&  CS97			 \\%check uncertainty
$V$                  	&  12.77 $\pm$ 0.01    		& HM01, HM08	 \\
$V$                  	&  12.65 $\pm$ 0.19    		& APASS            	\\
$V$                  	&  12.81 $\pm$ 0.11    		& CMC           	\\
$V$                  	&  12.79 $\pm$ 0.07    		& ASAS  		\\
$V$                  	&  12.79 $\pm$ 0.13    		& TASS IV            \\
$V/R$	        	&  12.82 $\pm$ 0.12    		& NSVS           	 \\
$r'$                 	&  12.44 $\pm$ 0.18$^a$    	& APASS           	 \\
%$r'$                  	&  12.58 $\pm$ 0.01$^a$		& CMC14           	 \\
$r'$                  	&  12.58 $\pm$ 0.01$^a$		& CMC15           	 \\
$R_{F}$       		&  12.38	$\pm$ 0.01			& ACR99		\\   %check
$R_{C}$       		&  12.31	$\pm$ 0.03			& HM01, HM08	 \\  %check
%$R_{F}$       	&  11.52						& GSC 2.3		\\
$i'$     	          	&  12.20 $\pm$ 0.11$^a$~~~~~~~~ & APASS          \\
$I_{C}$           	&  12.02 $\pm$ 0.07	     		& TASS IV		  \\						
$I_{C}$           	&  11.84 $\pm$ 0.05       		& HM01, HM08	 \\
$I_{g}$             	&  11.86 $\pm$ 0.02       		& DENIS        	 \\						
%$I_{N}$       	&  11.72						& GSC 2.3		\\
\hline
$J$                 	&  11.20 $\pm$ 0.06       		& DENIS    		 \\		
$J$                 	&  11.16 $\pm$ 0.03	  		& 2MASS     		\\		
$H$                	&  10.81 $\pm$ 0.02	   		& 2MASS      		\\						
$K_{s}$         	&  10.74 $\pm$ 0.07     		& DENIS   		\\
$K_{s}$         	&  10.74 $\pm$ 0.02	   		& 2MASS     		\\  
%\hline
$[$3.4]			&  10.65 $\pm$ 0.02	   		&  WISE     		\\
$[$4.6]		    	&  10.69 $\pm$ 0.02			&  WISE     		\\
\hline
$(U-B)_0$          	&   ~~~~$-0.07$				&  This work		\\%CHECK
$(B-V)_0$           	&   ~~~~$+0.52$				& This work		\\%
$(V-I_{C})_0$      	&   ~~~~$+0.62$				&  This work		\\%
~~~$(V-J)_0$~~~ &   ~~~~$+1.06$				& This work		\\%
$(V-K_{s})_0$   	&   ~~~~$+1.34$				&  This work		\\%
\hline
\end{tabular}
\end{center}
}
\begin{flushleft}
{\scriptsize $^a$AB magnitude.  References for photometry: 2MASS -- Skrutskie et al. (2006);  ACR99 --  Stone, Pier \& Monet (1999); APASS -- Henden et al. (2012);  ASAS -- Pojmanski (1997); CS97 -- Cieslinski et al. (1997); CMC --  Copenhagen University Obs. et al. (1999); CMC15 --  Niels Bohr Institute et al. (2011); DENIS  -- Epchtein et al. (1997), DENIS Consortium (2005); HM01 -- Henden \& Munari (2001); HM08 -- Henden \& Munari (2008); NSVS -- Wo\'zniak et al. (2004); TASS IV -- Richmond (2007); WISE -- Cutri et al. (2013). }
\end{flushleft}
\end{table}

%stellar colours, DeM+13, SK82, Cox
%F8 V, B-V = 0.53, Mv = +4.01, U-B = 0.02, V-I = 0.60, V-J = 1.06, V-K = 1.29
%F8 III, B-V = 0.54, Mv = +1.3, U-B = 0.41,
%F8 II, B-V = 0.58, Mv = -2.3
%F8 Ib, B-V = 0.56, Mv = -5.1, U-B = 0.41, V-I = 0.67, V-K = 1.37

%99 Her (F8 V), V = 5.04, B-V = 0.52, U-B = -0.10
%12 Boo (F8 V), V = 4.83, B-V = 0.54, U-B = 0.07, V-Ks = 1.23
%94 Cet (F8.5 V), V = 5.06, B-V = 0.56, U-B = 0.12, V-Ks = 1.31
%omi Aql (F8 V), V = 5.12, B-V = 0.55, U-B = 0.08, V-Ks = 1.22
%HR 244 (F8 V), V = 4.80, B-V = 0.54, U-B = 0.09, V-Ks = 1.16
%ups Peg (F8 III), V = 4.41, B-V = 0.61, U-B = 0.14, V-Ks = 1.38
%Polaris (F8 Ib), V = 2.02, B-V = 0.60, U-B = 0.38, V-Ks = 1.56 

\begin{figure*}
\begin{center}
\includegraphics[width=15.cm]{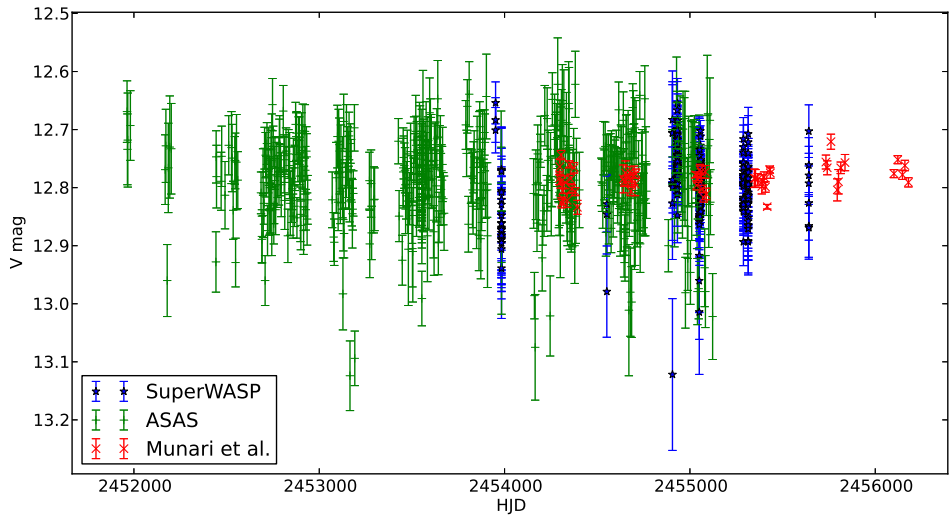}
\caption{Light-curve for DT\,Ser, based on photometric data from MC13 and the ASAS and SuperWASP surveys (refer to the text).  All three data sets were normalised to the weighted average magnitude of the ASAS $V$-band data. }
\label{DTSer_lightcurve}
\end{center}
\end{figure*}

\begin{figure}
\begin{center}
\includegraphics[width=8.2cm]{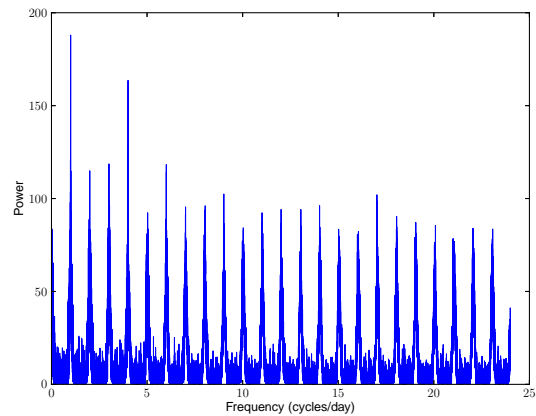}
\caption{Lomb-Scargle periodogram for DT\,Ser, showing peaks located at one cycle per day and every multiple of one hour periods.}
\label{DTSer_ls}
\end{center}
\end{figure}

\begin{figure}
\begin{center}
\includegraphics[width=8.2cm]{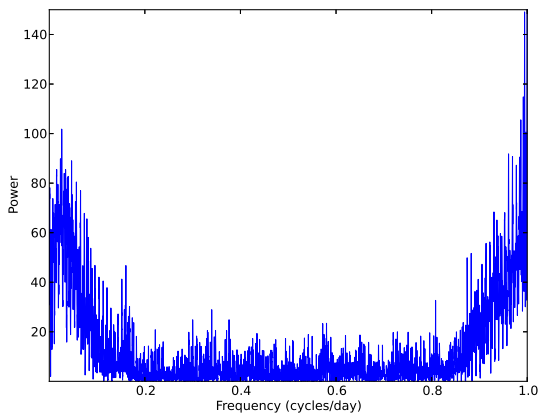}
\caption{Periodogram showing periods less than one cycle per day. An increase in significance is observed for long periods.}
\label{DTSer_ls_zoom}
\end{center}
\end{figure}

To detect any variability from DT\,Ser, and to further assess its classification as a putative symbiotic star, we combined the $V$-band photometric data in MC13 with $V$-band magnitudes from  the All Sky Automated Survey\footnote{\url{http://www.astrouw.edu.pl/asas/?page=aasc}} (ASAS; Pojmanski 1997), and data from the SuperWASP survey (Pollacco et al. 2006), which cover an 11 year period in total.  The largest data set of 403 points comes from ASAS, which gives a mean $V$-band magnitude of $12.785\pm0.069$ (1-sigma dispersion).  We have normalised all three data sets to the ASAS mean magnitude, and the resulting light-curve is presented in Figure\,\ref{DTSer_lightcurve}.  
Due to the non-uniform data sampling we have adopted a Lomb-Scargle periodicity analysis, and tested for periods as low as one hour. The resulting periodogram, shown in Figure\,\ref{DTSer_ls}, exhibits peaks 
at one cycle per day and every multiple of one hour periods. This is a consequence of the sampling frequency of the ASAS data and the one day cycle of long-baseline observations and is a natural artefact of the Lomb-Scargle method. The analysis is therefore consistent with no periodic variations for DT\,Ser.  All tested periods over one day (less than one cycle per day) are highlighted in Figure\,\ref{DTSer_ls_zoom}. 
An increase in significance is observed for long periods, which is potentially indicative of some residual systematic noise in the light curves, but entirely consistent with no detected periods of astrophysical nature.

%Besides the 2MASS data reported in MC13, we also used  near-IR data from the All WISE Data Release (Cutri et al. 2013).   
%Identifications of Sonneberg variables (Kinnunen+, 1999-2000) 

We also downloaded the ASAS data for all 615 stars detected within 30\arcmin\ of DT\,Ser, investigating the root-mean-square (RMS) scatter as a function of $V$ magnitude for each individual star (each has between 50 and 420 independent points). Based on this analysis, DT\,Ser would be considered constant in brightness down to an amplitude of 0.07 mag, which happens to be the average RMS variation at this magnitude level.  This conclusion is consistent with examination of the less precise photometry from The Amateur Sky Survey\footnote{\url{http://stupendous.rit.edu/tass/tass.shtml}} (TASS-IV; Droege et al. 2006; Richmond 2007), and the AAVSO Photometric All-Sky Survey\footnote{\url{http://www.aavso.org/apass/}}  (APASS; Henden et al. 2012).  In addition, $J$ and $K_s$ magnitudes are available from two different  surveys, DENIS and 2MASS (Epchtein et al. 1997; Skrutskie et al. 2006), and the magnitudes agree within the uncertainties.  The low-precision photographic measurements compiled by Jurdana-Sepic \& Munari (2010) also imply constancy within the uncertainties.  In summary, we conservatively estimate that any intrinsic variability is less than 0.1\,mag, much lower than the photographic range given in the GCVS, which we consider is spurious.

In summary we are unable to detect any periodic behaviour on a time scale of several months to years symptomatic of a symbiotic star.  Our optical light-curve also shows no evidence for either eclipses or eruptive events.  While the light-curve cannot definitively exclude a symbiotic interpretation as the system could be at an unfavourable inclination, this interpretation is highly unlikely once all of the other observational evidence is taken into account.

With a spectral type of F8, an inferred surface temperature of $\sim$6100\,K (Cox 2000), and the assumption that the star is on, or near, the main sequence, DT\,Ser is located in a part of the Hertzsprung-Russell Diagram where intrinsic variability is uncommon (e.g. Eyer \& Mowlavi 2008). DT\,Ser is noticeably cooler than the classical instability strip at this luminosity, so pulsational variability is not expected.  A perusal of the literature confirms this, with the $\delta$\,Scuti and $\gamma$\,Doradus stars (e.g. Uytterhoeven et al. 2011) being considerably hotter than DT\,Ser.   However, if DT\,Ser is shown in the future to have low-level variability, it may either be due to chromospheric activity, or the star is possibly part of a grazing eclipsing binary system.  However, further investigations are needed to definitively show if DT\,Ser should be reclassified as a constant star.\footnote{Many named variable stars in the latest edition of the GCVS are denoted as CST (constant).  Two brighter examples are Z\,Gem (Parkhurst 1905) and X\,Tau (Gaposchkin 1952); see Schmidt (1996) for some other cases.} 

%On the assumption that the star is on, or near, the main sequence we would not expect any obvious intrinsic variability at this spectral type.  The star is noticeably too cool to be a $\delta$ Scuti pulsator.  On the other hand, the star could be a member of an eclipsing binary, but we have found no evidence for this.

But how do we explain the large magnitude range detected on the Sonneberg plates?   Without having access to the original plates we can hypothesise that variations in  sky brightness, atmospheric transparency, exposure time and the technique used to measure the plates would cause the PN image to vary in its detectability.   Thus its variable contribution to the blended image with DT\,Ser on the old patrol plates could easily explain the range in photographic brightness recorded by the Sonneberg observers.
%What is the instrument at Sonneberg?

%MHZ give $V$ = 16.23 and $B-V$ = +0.37 for star\,B.

\subsection{Proper Motion}
In addition, DT\,Ser has a significant proper motion of +28.5, +31.0\,mas\,yr$^{-1}$ according to the PPMXL astrometric catalog (Roeser et al.  2010), though this is inconsistent with the UCAC4 proper motion of +9.3, +14.9 mas\,yr$^{-1}$ from Zacharias et al. (2012).  If real, the magnitudes of the proper motions add further weight to the conclusion that DT\,Ser is near the main sequence, though the blended image on most archival photographic plates suggests the inconsistent proper motions may be spurious.

\subsection{Reddening and Distance}\label{sec.DT_distance}

%The spectrum of DT\,Ser is consistent with it being a main sequence star, though a giant cannot be definitively ruled out. They calculated distances for luminosity classes of V, III and Ib respectively, after adopting a colour excess of $E(B-V)$ = 0.19 mag. The spectroscopic distance of 600\,pc is consistent with  some reddening from dust clouds of the Aquila Rift (Strai\v{z}ys et al. 2003). 

%The spectrum of DT\,Ser is consistent with it being a main sequence star, though a giant cannot be definitively ruled out.  

From the optical and infrared magnitudes, MC13 derived a modest reddening to DT\,Ser of $E(B-V)=0.19$ using the intrinsic colours of an F8 star.  We have recalculated the reddening using all available literature photometry (Table\,\ref{table:phot}) and adopting the intrinsic colours from De Marco et al. (2013), finding a slightly higher value of $E(B-V)=0.25$.
While we prefer a main-sequence classification for DT\,Ser, we will quote a range of III-V for the luminosity class.  Using our reddening value, the distance to the star is between 400 and 1400\,pc for absolute magnitudes of +4.0 (F8 V) and +1.3 (F8 III) respectively (De Marco et al. 2013; Schmidt-Kaler 1982).  

% Perhaps save this till later? - no need to make the point twice
%We will see in \S\,\ref{sec:PN_distance} that DT\,Ser is much closer than the distance we estimate to the PN, providing very strong evidence against any association between the two.

The stellar reddening is only about half of the total reddening on this sightline from the revised dust maps of Schlafly \& Finkbeiner (2011), which is $E(B-V)=0.51$.  %Furthermore, a supergiant classification cannot be reconciled with the relatively low extinction.
Blue and red SSS images show that patchy yet significant extinction can be expected from the Serpens Cauda dark cloud complex (Strai\v{z}ys, Barta\v{s}\={u}t\.{e} \& Cerni\v{s} 2002), part of the Aquila rift in this direction.  Strai\v{z}ys, Cerni\v{s} \& Barta\v{s}\={u}t\.{e} (2003) determined the distance to the near side of the rift to be $225\pm55$ pc with a thickness of $\sim$80\,pc, producing $V$-band extinctions of a few tenths up to 3.0\,mag.  The observed reddening to DT\,Ser suggests that this sightline is less opaque than the densest parts of the rift (confirmed from SSS images).   The additional reddening to the PN (\S\,\ref{sec:reddening}) is the result of interstellar material lying beyond the nearby Serpens clouds.

%%%%%%%%%%%%%%%%%%%%%%%%%%%%%%%%%%%%%%%%%%%%
\section{The Planetary Nebula}\label{sec:PN}
%%%%%%%%%%%%%%%%%%%%%%%%%%%%%%%%%%%%%%%%%%%%

\subsection{Nebular Morphology}

\begin{figure*}
\begin{center}
\includegraphics[width=17cm]{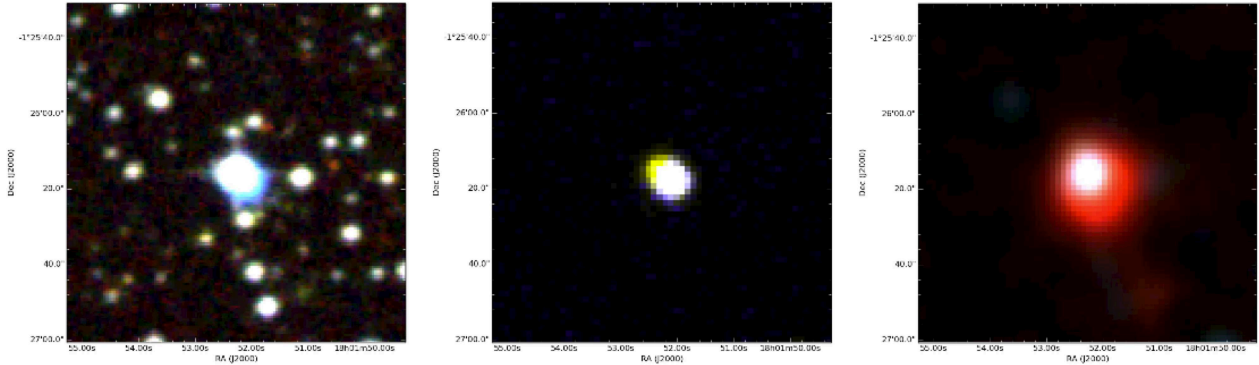}
\caption{Montage of images showing DT Ser and the unrelated planetary nebula MCS\,1, all at the same scale. In order, we show a SuperCOSMOS $B_{J}$/$R_{F}$/$I_{N}$ composite, a GALEX FUV/NUV false colour composite, and a WISE 3.4/4.5/12\,$\mu$m composite.  The bluish PN is just visible in the optical image, close south-west of DT\,Ser, which has diffraction spikes. Two slightly offset components are seen in the combined GALEX image, with the hotter (bluish) component at lower left representing the CSPN. The WISE image clearly shows DT\,Ser as a white compact component offset from the centroid of the longer wavelength 12\micron\ feature which represents the PN.   Each panel is 90\arcsec on a side with NE at top-left.
A colour version of this figure is available in the online journal.   }
\label{fig.montage}
\end{center}
\end{figure*}

MC13 presented high-resolution \ha\  and \OIII\ images of MCS\,1 taken in good seeing with the 2.6-m Nordic Optical Telescope (NOT).  The images show a double-shelled nebula with a faint round outer shell and an elliptical inner rim.  The overall angular diameter is 11\farcs4.  In Figure\,\ref{fig.montage}, we present a montage showing GALEX ultraviolet, SSS optical, and WISE infrared images of DT\,Ser and its superposed PN.  
While MC13 noted that the apparently distorted inner rim of the PN shell suggested an association with Star\,A, we find this argument non-compelling. 

It is true that their \OIII\ image in particular appears to show that the inner ring is elongated towards star A. However, this could simply reflect the ellipticity of the inner PN shell with a major axis position angle aligned coincidentally with star A.   The slightly off-centred central star is not an issue, as many CSPNe are not located at the exact geometric centres of their nebulae (e.g. Sahai et al. 1999; Bobrowsky et al. 2004; Pereira et al. 2008).  Such an effect is possibly due to the central star being part of a wide binary system (Soker 1994; Soker, Rappaport \& Harpaz 1998) or alternatively is the product of the PN's interaction with the ISM (e.g. Wareing et al. 2006).

\subsection{Integrated Fluxes}\label{nebular_fluxes}   
MC13 give a \ha\ flux of 1.55 $\times$ 10$^{-13}$ \ergcms\ derived from their long-slit spectrum.  This is a lower limit to the true flux, as (i) the slit width did not include all of the PN, and (ii) the \ha\ absorption feature in the spectrum of the F8 star may have affected the nebular flux.  By scaling up the measured flux to the full area of the PN, we obtain a crude estimate of the total flux as 9 $\pm$ 3 $\times$ 10$^{-13}$ \ergcms.   Fortunately,  the nebula is visible on SHASSA \ha\ images (Gaustad et al. 2001), so we can determine an independent integrated \ha\ flux following the recipe of Frew, Boji\v{c}i\'c \& Parker (2013).  Our determination, $F$(\ha) = 7.1 $\pm$ 1.0 $\times$ 10$^{-13}$ \ergcms, is consistent with the scaled-up estimate. 
There is also a radio detection of the PN in the 1.4\,GHz NRAO VLA Sky Survey (NVSS; Condon et al. 1998), designated NVSS 180152-012623, with a total flux of $\sim$4.3\,mJy.  The NVSS position is 8\arcsec\ SE of Star A, and 4\arcsec\ SSE of Star B, which corresponds to the planetary nebula, given the typical NVSS astrometric uncertainty of $\sim$3\arcsec\ for a source with this flux (see Fig. 30 of Condon et al. 1998).

We can also estimate an approximate integrated \OIII\ flux from the available literature data.  While Cieslinski et al. (1997) caution against using the fluxes they measured owing to continuum contamination, the graphical plot in their Fig.\,1 can be used to estimate a rough $\lambda$4959/$\lambda$4861 intensity ratio of $\sim$4.5 for the PN, with an error of about 50 per cent (note the \ha\ flux cannot be estimated directly from their plot).  Since at the adopted reddening (see below) the $\lambda$5007/$\lambda$4959 and \ha/\hb\ line ratios are predicted to be 3.0 and 4.9 respectively, we can infer that the raw 5007/\ha\ ratio must be $\sim$3.1, quite typical for a high-excitation PN (Ciardullo 2010).  From the measured \ha\ flux, an approximate integrated \OIII\ flux, $F$(5007) = 2.2 $\pm$ 1.0 $\times$ 10$^{-12}$ \ergcms\ is determined.  Using the formula of Jacoby (1989), this flux corresponds to an apparent $V$-band equivalent magnitude for the PN of +15.4, with an uncertainty of perhaps half a magnitude.

%total app. mag of PN + CSPN = 15.0;  V_5007(0) = +13.6

\subsection{The Central Star}\label{sec:CSPN}
The central star is clearly visible in the 2.6-m NOT images reproduced in MC13.  Photoelectric $UBVRI$ photometry of this star was presented by Cieslinski et al. (1997), and more recent CCD PSF-fitting photometry was given by Henden \& Munari (2008).  The CCD data, taken at three epochs, show the CSPN to be constant within the uncertainties, with a mean magnitude of $V$ = 16.21.  We also locate photometry from GALEX on the AB system (Bianchi et al. 2011).  Table\,\ref{table:cspn} summarises all literature photometry which unambiguously refers to the central star.

%Table\,\ref{table:cspn} summarises the optical photometry of the CSPN from MC13, while the GALEX ultraviolet magnitudes were adopted from Bianchi et al. (2011).
%$m_{\rm FUV} =16.58\pm0.03$ and  $m_{\rm FUV} = 17.06\pm0.03$ on the AB system (Bianchi et al. 2011).

\begin{table}
{\footnotesize	
\begin{center}
\caption{Summary of photometric measurements for the PN central star.  The dereddened colour indices were calculated assuming $E(B-V)$ = 0.50.}
\label{table:cspn}
\begin{tabular}{cll}
\hline
Waveband~ 				&    ~~~~~~~~mag 					&  Source				\\ %
% 						&      								&  						\\ %$A_{V} = 1.55$ check
\hline							
$FUV$                 			&  16.58 $\pm$ 0.03$^a$				&  GALEX				\\ %
$NUV$                			&  17.06 $\pm$ 0.03$^{a\dag}$		&  GALEX				\\ %
%\hline
$U$                  			&    15.32 $\pm$ 0.07~~~~~~			&  CS97					\\ %
$U$                  			&    15.76 $\pm$ 0.03					&  HM08					\\ % 15.764 $\pm$ 0.03; delta = 0.44
$B$                  			&   15.91 $\pm$ 0.07					&  CS97					\\ %
$B$                  			&   16.58 $\pm$ 0.06					&  HM08					\\ %16.584 $\pm$ 0.06; delta = 0.67
$V$                  			&   15.40 $\pm$ 0.05					&  CS97					\\ %
$V$                  			&   16.21 $\pm$ 0.05					&  HM08					\\ % 16.213 $\pm$ 0.05; delta = 0.81
$R_{c}$       				&   15.34 $\pm$ 0.06	 				&  CS97					\\ %
$R_{c}$       				&   16.27 $\pm$ 0.06 					&  HM08					\\ %16.268 $\pm$ 0.06; delta = 0.93
$I_{c}$           			&   15.44  $\pm$ 0.07					&  CS97					\\ %
$I_{c}$           			&   16.26 $\pm$ 0.05  				&  HM08					\\ %16.257 $\pm$ 0.05; delta = 0.82
\hline
~~$(F-V)_0$~~                  &	~~~~~$-2.3:$  					&  This work				\\ %
$(U-B)_0$                  		&	~~~~~$-1.18$ 	 				&  This work				\\ %
$(B-V)_0$                  		&	~~~~~$-0.13$					&  This work				\\ %
%$(V-R_c)_0$                	&	~~~~~$-0.34$					&  This work				\\ %
%$(R_c-I_c)_0$                	&	~~~~~$-0.34$					&  This work				\\ %
$(V-I_c)_0$                		&	~~~~~$-0.68$					&  This work				\\ %
\hline
\end{tabular}
\end{center}
}
{\footnotesize $^a$AB magnitude; $^{\dag}$flux dominated by F8 star.  References for photometry: CS97 -- Cieslinski et al. (1997);  HM08 -- Henden \& Munari (2008); GALEX -- Bianchi et al. (2011)}
\end{table}

%HM08: V = 16.213, U-B = -0.82, B-V = 0.371, V-R = -0.055, R-I = 0.011 (0.11 in MC13), V-I = -0.044 (+0.055)
%CS97: V = 15.40, U-B = -0.51,  B-V = 0.51, V-R = -0.06, R-I = 0.10, V-I = 0.04

In their paper, MC13 suggested the 0.8\,mag difference between the photoelectric photometry of Cieslinski et al. (1997) and the CCD photometry of Henden \& Munari (2008) provided evidence for intrinsic variability.  Given how close the stars are together and how star B is about 3.4 mag fainter than star A, we believe that photometric contamination from the nebula, and potentially the brighter star as well, has affected the earlier results.  Recall that the PN's apparent magnitude, $V$ = 15.4 (from \S\,\ref{nebular_fluxes}), is brighter than the central star itself, and some flux from the PN would probably have been included in the photoelectric measurement. In summary, there is currently no observational evidence for variability of the CSPN, though dedicated time-series photometry is needed to make a definitive statement. 

We note in passing that a few bona fide old PNe also have central stars with superposed \textit{high-density} forbidden emission lines, which Frew \& Parker (2010) called EGB\,6-like central stars, after the archetype EGB\,6 (Liebert et al. 2013).  The 2-D Echelle spectrum across the central star of MCS\,1 (reproduced in Fig. 1 of MC13) shows it is unlikely to be a member of this group (see Miszalski et al. 2011a,b, for some other examples of the class), nor is it a likely emission-line CSPN (e.g. DePew et al. 2011).

\subsection{Reddening}\label{sec:reddening}

We estimate the reddening to MCS\,1 using three separate methods. As noted by MC13, the reddening estimated from the photometry of the F8 star cannot reproduce the colours of the central star\,B for any spectral type.  Using the photometry from Table\,\ref{table:cspn}  Assuming the CSPN's effective temperature is \grtsim60\,kK, necessary to explain the nebular \HeII\ emission, we use the photometry from Table\,\ref{table:cspn} and adopt the intrinsic colours of a 60\,kK blackbody from De Marco et al. (2013) to estimate $E(B-V)$ = 0.50 $\pm$ 0.10.  Comparing the integrated \ha\ and NVSS 1.4\,GHz fluxes leads to an second, independent determination of the reddening to the PN, $E(B-V)$ = 0.7 $\pm$ 0.2, following Boji\v{c}i\'c et al. (2011a,b).  The considerable uncertainty results from the uncertainties in both the \ha\ and radio fluxes.  
Never the less, both these estimates agree with the asymptotic reddening on this sightline, $E(B-V)$ = 0.51, derived from the revised reddening maps of Schlafly \& Finkbeiner (2011; SF11).  There is therefore no strong evidence for any  internal extinction due to the PN shell, which would be unlikely given the evolved nature of the PN (see the discussion by Giammanco et al. 2011).  We will adopt the SF11 reddening from now on.
  
Other objects in the vicinity lend support to a higher value for the {\it total} reddening in this direction than assumed by MC13. For the extreme-He star BD\,$-$1\arcdeg3438 (NO\,Ser), 40\arcmin\ north-east of DT\,Ser,  Heber \& Sch\"onberner (1981) estimate a visual absorption of 1.76 mag, or  $E(B-V)=0.57$, for a distance of 4.7\,kpc.  This agrees within the uncertainties with the SF11 value, $E(B-V)=0.51$, and shows the value adopted by MC13 was too low.

%Now considering the scale-height of the dust layer (cf. MC13), and the fact that the extinction to the star is less than half of the total extinction along this sightline (Schlafly \& Finkbeiner 2011), we conclude the distance is less than 1.0\,kpc.

\subsection{Distance and Derived Properties}\label{sec:PN_distance}

Simple algebra shows a PN as small and as faint as this one must be well beyond the 1.0\,kpc volume-limited sample of Frew (2008).  We can use our integrated \ha\ flux, the nebular reddening and the angular diameter of 11.4\arcsec\ from MC13 to determine the distance to the PN using the \ha\ surface brightness -- radius ($S$--$r$) relation of Frew (2008) and Frew et al. (2014).  The observed and reddening-corrected logarithmic \ha\ surface brightnesses are log\,$S$(\ha) = $-3.52$ and log\,$S_0$(\ha) =  $-3.02$ \ergcms\,sr$^{-1}$ respectively.  Since the PN is  moderately evolved and of high excitation (i.e. weak \NII\ and moderate \HeII\ emission lines), it is likely to be optically thin, so the `short' trend relation is applicable, from which we estimate a linear radius of 0.18\,pc.  Using the observed angular size the distance is readily calculated to be $6.4\pm2.0$\,kpc.  This distance places MCS\,1 behind the F8\,III-V star DT\,Ser.   The absolute \OIII\ $\lambda$5007 magnitude is $-0.4$, after correcting for extinction.  This  is $\sim$4.1 magnitudes down from the tip of the \OIII\ PN luminosity function (PNLF; Ciardullo 2010), consistent with the relatively evolved nature of this optically thin PN.

At a Galactic latitude of +10.3\arcdeg, the PN is located more than 1.1\,kpc from the plane, suggesting it is an old thin-disk or thick-disk object.  We note that if the optically-thick $S$--$r$ relation from Frew (2008) is applied, then the distance is $\geq$8\,kpc, but we consider the shorter distance the more likely, based on the spectroscopic characteristics of the PN.   Round PNe are known to be at larger average distances from the Galactic plane than other morphological types, either inferred from the Galactic latitude distribution (Phillips 2001; Parker et al. 2006; Jacoby et al. 2010) or determined directly from volume-limited samples (Frew 2008).  Many round PNe are also optically thin (e.g. Kaler 1981; Jacoby, Ferland \& Korista 2001), so in these respects, MCS\,1 is a typical exemplar.

From the nebular flux and our preferred distance of 6.4\,kpc, we calculate a root-mean-square electron density, $n_e$ = 320\,cm$^{-3}$, and an ionized mass of 0.16\,$M_{\odot}$ (for a filling factor of unity), typical values for a middle-aged PN of relatively low mass.  Using the expansion velocity of 37\,\kms\ from the \OIII\ line splitting (MC13), we estimate an approximate dynamical age of 4700 years for the PN shell.   In Table\,\ref{table:dist_comparison} we give a comparison of the properties of the PN at our adopted distance, and at the distance we have estimated for DT\,Ser.  A comparison of these properties with other PNe (see Frew \& Parker 2010), in particular the ionized mass and  \OIII\ absolute magnitude, show that the shorter distance estimates can be definitively ruled out. 

\begin{table}
\begin{center}
\caption{A comparison of the physical properties of Star\,A and the PN, at near, intermediate, and far distance estimates (refer to the text).}\label{table:dist_comparison}
\begin{tabular}{l c c c}
\hline
Distance (pc)~~~~~~			& 		400 						& 	1400						&    6400			 \\ 
\hline
$M_V^{\rm A}$ 				& 		+4.0						& 	+1.3							& 	$-2.0$	 		\\
%$L_{\rm A}$ ($L_{\odot}$)	& 		2.0						&								& 		 			\\
%$L_{\rm B}$ ($L_{\odot}$)	&								&  								&	 				\\
$z$ (pc)						&		72						& 	250							&	1150 			\\
$m_{\rm ion}$  ($M_{\odot}$)	& 1.2\,$\times10^{-4}$			&~~~~~2.7\,$\times10^{-3}$~~~~~	&	0.16 			\\
$r_{\rm shell}$  (pc)			&  		0.011					& 	0.039						&	0.18				\\
$T_{\rm kin}$ (yr)				&		290						& 	1000						&	4700 			\\
$M_{5007}$					&		+6.5						& 	+3.8							&	~~~$-0.4$~~~ 	\\
\hline
\end{tabular}
\end{center}
\end{table}

\subsection{The Mid-infrared Source}\label{sec.MIR}

DT\,Ser is ostensibly an IRAS mid-infrared (MIR) source, designated IRAS\,17592-0126, with 25 and 60\,$\mu$m fluxes of 299\,mJy and 608\,mJy respectively. However, the coarse resolution of the IRAS survey means it is impossible to tell if the MIR source is coincident with the brighter star, the PN, or both.  In addition, the WISE survey has a detection in all four bands, with the astrometry showing that the WISE band-1 (W1) and W2 magnitudes unambiguously refer to star A.  However MC13 assumed that the WISE W3 and W4 detections (and by extension the IRAS fluxes) also apply to this star.  Thus MC13 claimed that DT\,Ser had a large MIR excess and was possibly a post-AGB star, using this interpretation to potentially explain the observed velocity shift between star\,A and MCS\,1.  Recall that we showed in \S\,\ref{sec:spectrum} that DT\,Ser cannot be a low-gravity post-AGB star on the basis of its spectrum and $(U-B)_0$ colour.

We carefully examined the WISE images and found that the centroid of the WISE W3 and W4 point-spread function (PSF) is clearly offset from the W1 and W2 PSFs by about 5\arcsec\ and is within 1\arcsec\ of the nominal PN position (see Figure\,\ref{fig.montage}).  Our analysis shows that the W1 and W2 magnitudes clearly refer to star\,A, while the W3, W4, and IRAS fluxes to refer to MCS\,1.   The MIR colours are similar to other PNe (Anderson et al. 2012; Parker et al. 2012), but since the PN is optically thin, with moderately strong \HeII\ emission, we would expect that much of the W4 flux is due to the \OIV\ fine-structure line at 25.9\micron\ (e.g. Chu et al. 2009; Fesen \& Milisavljevic 2010), with only a small contribution from warm dust.  
%In Figure\,\ref{fig:WISE} we plot the \ha\ surface brightness against the 22\micron\ surface brightness for 60 optically thin PNe (data primarily taken from Frew 2008). MCS\,1 is seen to be a very typical PN in its properties. 

\subsection{Is the nebula variable?}\label{sec.PN_var}

The conflicting information in the literature about MCS\,1 and its lack of detection by earlier authors led MC13 to speculate that the PN has varied with time.  This would be an almost unique occurrence for an {\it evolved} PN, with the only roughly comparable examples known to the authors being the flash-ionized PN around the classical nova V458\,Vul (Wesson et al. 2008) and the slow secular variation seen in the old PN Hen\,1-5 (Henize 1969; Arkhipova, Esipov \& Ikonnikova 2009) around the born-again star FG\,Sge.  However there is no evidence to suggest an eruptive event in either star\,A or B has ever been seen on archival plates, even though Henden \& Munari (2001) and MC13 speculated that star\,B may be strongly variable, perhaps based in part on the mis-classification by Cieslinski et al. (1997) that star\,B is a symbiotic star.  

The surface brightness of MCS\,1 is modest, and it may have been simply missed by Bond (1978), though it is unclear if his observation ``at the telescope'' was through the eyepiece or noted from an acquisition screen of some sort.  
We also examined all available blue and red Schmidt plate scans from the MAST DSS and SSS websites.  %ADD URLs
The PN clearly contributes to the image blend on all three blue and both red plates taken with the Palomar and UK Schmidt Telescopes between 1953 and 1990.  Hence there is no evidence for any strong PN variability based on this data set.  Such a conclusion is unsurprising considering we have determined a very different density for MCS\,1 ($n_e$\,=\,320\,cm$^{-3}$), based on its \ha\ surface brightness, compared to the high density calculated by MC13.

%see the following blue and red plates (plate no., emulsion/filter, date):\\
%\noindent{POSS plate 773; 103aE; 1953-07-11}\\
%UKST plate 5045; IIIaJ+GG395; 1979-05-28\\
%UKST plate 12684; IIIaF+OG590; 1988-08-04\\
%POSSII plate SJ03190; IIIaJ+GG395; 1990-05-01\\
%POSSII plate SF03388; IIIaF+RG610; 1990-07-23\\

\subsection{A short comparison with symbiotic outflows}\label{sec:sys}
Frew \& Parker (2010) and MC13 reviewed a number of resolved bipolar nebulae surrounding some symbiotic stars, like BI\,Cru (Schwarz \& Corradi 1992; Corradi \& Schwarz 1993), Hen\,2-104 (Santander-Garc\'ia et al. 2008), and probably OH\,231.8+4.2 (Bujarrabal et al. 2002; Meakin et al. 2003; Vickers et al. 2014).  Closely related nebulae include Mz\,3 (e.g. Santander-Garc\'ia et al. 2004; Cohen et al. 2011), Hen\,2-25 and Th\,2-B (Corradi 1995), and M\,2-9 (Corradi et al. 2011, and references therein), which also overlap in their observational properties with some strongly bipolar PNe (see Corradi \& Schwarz 1995).  The very high expansion velocities seen in many of these objects suggest they were expelled in a transient event, possibly associated with the binary nature of their nuclei (e.g. Soker \& Kashi 2012).   However all of these objects are morphologically and kinematically distinct from the nebula MCS\,1.  Based on the results of this study, the number of classical symbiotic stars (Kenyon 1986) in elliptical or round PNe is zero.

\section{Summary}\label{sec:summary}
We have offered an alternative interpretation for the putative symbiotic star DT Serpentis, and its supposed planetary nebula MCS\,1, recently announced by MC13.  Our new  scenario is based on clarifying the past confusion as to which of the two stars is actually DT\,Ser, pointing to the true nature of the WISE detections of DT\,Ser and MCS\,1.  We used the data presented in their paper supplemented with publicly available data obtained from Virtual Observatory databases.  After applying Ockham's Razor, we make the following conclusions:
\begin{enumerate}
\item The designation DT\,Ser applies to the brighter 12.7-mag star superposed on the periphery of MCS\,1 (star\,A of MC13), and should not be applied to the PN's central star (star\,B).
\item DT Ser is a foreground F8\,III-V star 400 -- 1400 pc away.  The nearer distance is more likely.
\item We find no evidence for any periodic variability of DT\,Ser.  While MC13 find evidence of secular variability, an analysis of more than a decade of ASAS and SuperWASP data did not confirm this, down to a limit of $\pm$0.07 magnitude.  While the possibility remains the star is a low-level variable, it is not a symbiotic star.
\item There is currently no observational evidence for any variability of the CSPN, though dedicated time-series photometry is needed to make a definitive statement. 
\item The substantially different ($\sim$60\,\kms)  radial velocities of DT\,Ser and MCS\,1 appear to preclude any association, though we note the stellar velocity was only a single epoch measurement. 
\item The reddening of MCS\,1 determined from the colours of star\,B, and from a comparison of the \ha\ and NVSS 1.4\,GHz radio flux measurements are consistent, and agree with the asymptotic reddening on this sightline from Schlafly \& Finkbeiner (2011).  MCS\,1 is more heavily reddened than DT\,Ser, as expected for a greater distance.  
\item Based on the \ha\ S-$r$ relation we estimate a distance of 6.4\,$\pm$\,2.0\,kpc to the PN, placing it in the far background of DT\,Ser. 
\item The longer wavelength mid-infrared WISE (and IRAS) detections are unambiguously associated with the high excitation PN, and not DT\,Ser.  There is therefore no evidence that star\,A is a dusty post-AGB star, nor any evidence to suggest that the PN itself has varied markedly with time.
\end{enumerate}

\section*{Acknowledgements}

%We thank the anonymous referee for constructive comments that improved this paper. 
This research has made use of the SIMBAD database and the VizieR service, operated at CDS, Strasbourg, France, and the APASS database, located at the AAVSO web site. Funding for APASS has been provided by the Robert Martin Ayers Sciences Fund.  D.J.F. thanks Macquarie University for a MQ Research Fellowship. J.B. is currently funded by an Australian Research Council Discovery Project and I.S.B. is the recipient of an Australian Research Council Super Science Fellowship (project ID FS100100019).  Q.A.P acknowledges support from the Australian Astronomical Observatory.

\end{document}